\documentclass[10pt,onecolumn]{article}
\usepackage{amsmath}
\usepackage{amsfonts}
\usepackage{amssymb}

\usepackage[latin1]{inputenc}
\usepackage{latexsym} % vgl. Kopka, p121
\usepackage{graphicx}
\usepackage{psfrag}
\usepackage{exscale}
\usepackage[permil]{overpic}       % zum berlagern von TeX-Ausdrcken und EPS-Bildern
\usepackage[vflt]{floatflt}        % Text umflie§t Grafiken und Tabellen
\usepackage{times}
\usepackage{makeidx}               % Anlegen eines Registers
\usepackage{color}
\usepackage{framed,color}
\usepackage{hyperref}
\usepackage{wasysym}
\usepackage{rotating}

\makeindex   % Anlegen eines Registers

\input epsf % defines \epsfbox and supporting macros
\epsfverbosetrue % messages will show height and width of figures

   {\begin{list}{}{%           (vgl Kopka, p200f)
      \settowidth{\labelwidth}{\textbf{#1}}
      \setlength{\leftmargin}{\labelwidth}
         \addtolength{\leftmargin}{\labelsep}
      \setlength{\parsep}{0.5ex plus0.2ex minus0.2ex}
      \setlength{\itemsep}{0.3ex}
      }}
   {\end{list}}

\newcommand{\ze}[2]{#1\;{\rm#2}} % neue Version von \pg
\newfont{\tensy}{cmsy10}           % Zur Darstellung chem. Formeln                          

\newcommand{\grad}{^\circ} % Winkel in Grad                                        
                                        
\newcommand{\fm}[1] % Formel
{\begin{displaymath}#1\end{displaymath}}

\newcommand{\nfm}[1] % numerierte Formel
{\begin{equation}#1\end{equation}}

\newcommand{\nmzfm}[1] % numerierte mehrzeilige Formel
{\begin{eqnarray}#1\end{eqnarray}}

\newcommand{\mzfm}[1] % mehrzeilige Formel
{\begin{eqnarray*}#1\end{eqnarray*}}

\newcommand{\gfm}[1] % gerahmte Formel
{\begin{displaymath}\fbox{$\displaystyle #1$}\end{displaymath}}

\newcommand{\ngfm}[1] % gerahmte und numerierte Formel
{\begin{equation}\fbox{$\displaystyle #1$}\end{equation}}

\newcommand{\mzgfm}[1] % mehrzeilige gerahmte Formel
{\begin{displaymath}\fbox{$\begin{array}{rcl} #1\end{array}$}\end{displaymath}}

\newcommand{\figcapt}[2] % genderte Bildunterschriften
{\renewcommand{\normalsize}{\small}\textbf{\caption[#1]{\textmd{#2}}}}

\newcommand{\fif}[1] % fett in Formeln
{\mathbf{#1}}
\newcommand{\bi}[4] % bestimmtes Integral
{\int_{#1}^{#2}{#3}\,{#4}}

\newcommand{\ubi}[2] % unbestimmtes Integral
{\int{#1}\,{#2}}

\definecolor{shadecolor}{gray}{.90} % Hintergrundfarbe (0.90 ist ein 
\newtheorem{beispiels}{Beispiel}[section] % Beispiel mit Abschnittnumerierung

\newtheorem{aufgabeS}{Aufgabe}[section]

\newcommand{\qM} % quadratische Matrix
{\left[\ddots\right]}

\newcommand{\SV} % Spaltenvektor
{\left[\vdots\right]}

\title{Numerical Investigation of Symmetry Breaking and Critical Behavior of the Acoustic Streaming Field in High-Intensity Discharge Lamps}

\author{\normalsize Bernd Baumann$^{*1}$,  Joerg Schwieger$^{1,2}$,  Marcus Wolff$^{1}$, \\ \normalsize Freddy Manders$^{3}$, and Jos Suijker$^{3,4}$}

\date{\footnotesize
$^{1}$Hamburg University of Applied Sciences,  Department of Mechanical Engineering and Production,\hfill \phantom{.}\\
\hspace{1mm} Berliner Tor 21, 20099 Hamburg,  Germany\hfill \phantom{.}\\
$^{2}$University of the West of Scotland, School of Engineering and Computing, High Street, \hfill \phantom{.}\\
\hspace{1mm} Paisley PA1 2BE, United Kingdom \hfill\phantom{.}\\
$^{3}$Philips Lighting, Steenweg op Gierle 417, 2300 Turnhout, Belgium \hfill\phantom{.}\\
$^{4}$Technical University Eindhoven, Den Dolech 2, 5612AZ Eindhoven, Netherlands \hfill\phantom{.}\\
$^*$ info@BerndBaumann.de\hfill\phantom{.}}

\begin{document}
\pdfoutput=1
\maketitle

\begin{abstract}
For energy efficiency and material cost reduction it is preferred to drive high-intensity discharge lamps at frequencies of approximately $\ze{300}{kHz}$. However, operating lamps at these high frequencies bears the risk of stimulating acoustic resonances inside the arc tube, which can result in low frequency light flicker and even lamp destruction. The acoustic streaming effect has been identified as the link between high frequency resonances and low frequency flicker. A highly coupled 3D multiphysics model has been set up to calculate the acoustic streaming velocity field inside the arc tube of high-intensity discharge lamps. It has been found that the velocity field suffers a phase transition to an asymmetrical state at a critical  acoustic streaming force. In certain respects the system behaves similar to a ferromagnet near the Curie point. It is discussed how the model allows to investigate the light flicker phenomenon. Concerning computer resources the procedure is considerably less demanding than a direct approach with a transient model.

\vspace{3mm}
\end{abstract}

\section{Introduction}
Worldwidely, 19\% of the electric power is consumed for lighting \cite{zissis2010diagnostics,waide2006light}. A considerable fraction of artificial light sources are high-intensity discharge (HID) lamps which are used for outdoor and shop lighting as well as automobile headlights and other applications. Despite an increasing market share of light emitting diodes, HID lamps will be irreplaceable in the foreseeable future because of their superior color rendering index and their sun like luminance \cite{NEMA2010}. Although HID lamps have reached a certain stage of technical maturity, further effort is required in order to obtain lamps of highest quality and efficiency.

\begin{figure*}[h]
\centering
\includegraphics[width=0.75\linewidth,trim=0 0 0 0,clip]{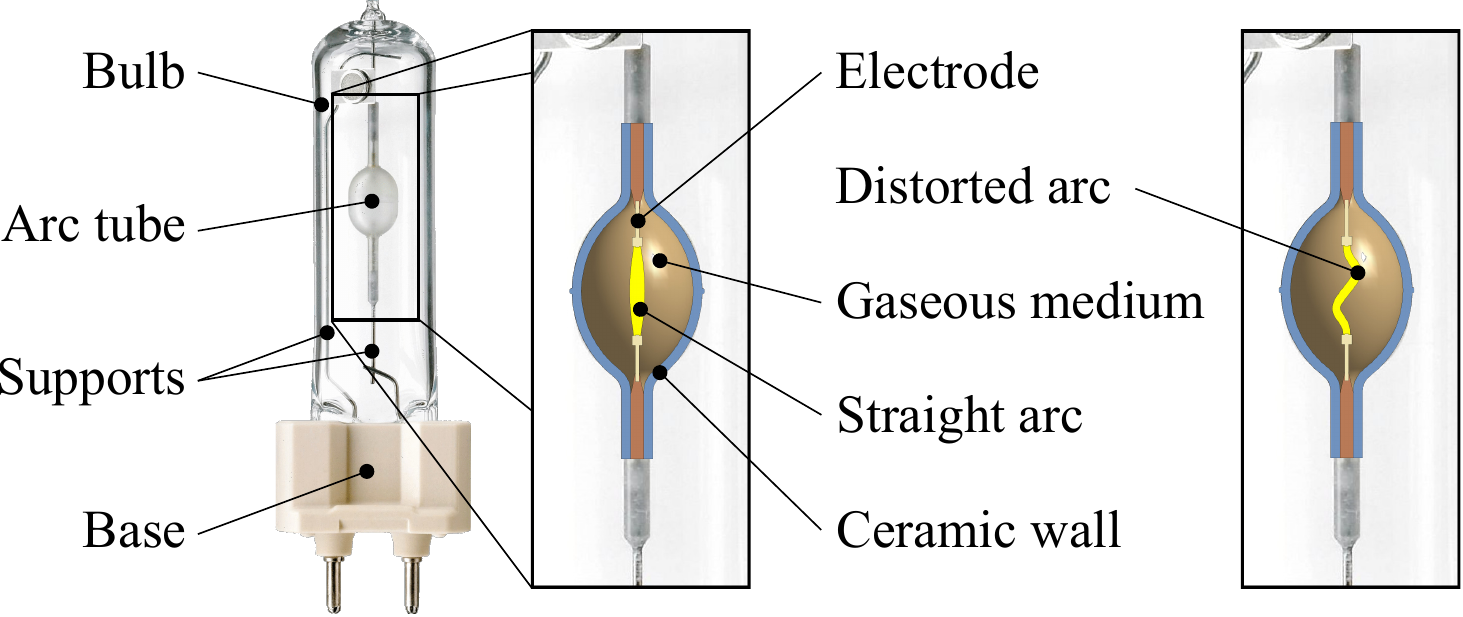}
\figcapt{}{Left: Design of the investigated HID lamp  (Philips 35 W 930 Elite). The arc tube is made from polycrystalline alumina (PCA) and contains mainly argon, mercury and metal halides. The distance between the tungsten electrodes is $\ze{4.8}{mm}$. Right: Arc perturbation in vertical lamp operation.\label{fig:StraightBowedArc_TNR}}
\end{figure*}

The design of the lamp investigated in this article is depicted in the left part of Figure \ref{fig:StraightBowedArc_TNR}. A voltage applied to the electrodes inside the arc tube establishes a plasma arc, which constitutes the source of light emission. To avoid electrode erosion and demixing of the arc tube filling, HID lamps are operated with alternating current (AC). From the electronic point of view minimal material costs are achieved at the energetically optimal operation frequency of approximately $\ze{300}{kHz}$ \cite{trestman2002minimizing}. Unfortunately, in this frequency range periodic heating due to ohmic loss excites acoustic resonances inside the arc tube. The high frequency sound wave causes low frequency fluctuations (ca. $\ze{10}{Hz}$) of the plasma arc that are visible as light flicker. It has recently been discovered that the acoustic streaming (AS) phenomenon is the link of the high frequency resonances to the low frequency light flicker \cite{afshar2008theory}. For further improvement of the lamp design a thorough understanding of the underlying mechanism is crucial. We present results for the calculation of the AS field obtained with a stationary 3D finite element model. As a test case we chose the  resonance of frequency $\ze{47.4}{kHz}$.

\section{Model}\label{sec:Model1}
\subsection{Temperature Field}
\label{sec:TemperatureField1}
The investigation of light flicker in HID lamps requires a model that comprises a number of coupled equations that describe the processes inside the arc tube and at the arc tube's wall and the electrodes. In contrast to the time-dependent 2D model that describes an arc tube of infinite length \cite{dreeben2008modelling}, we use a stationary 3D model. Arc flicker shall be identified through instabilities of the velocity field. %In principle, the velocity field is assumed to be a superposition of the buoyancy driven velocity field $\vec{u}$ and the AS velocity field $\vec{v}$. Since it turns out that the AS velocity is much larger than the buoyancy driven velocity, the latter will be neglected, i.e. we assume that the $\vec{v}$-field alone is responsible for arc flicker. 

\begin{figure}[htb]
\centering
\includegraphics[width=0.9\linewidth,trim=0 0 0 0,clip]{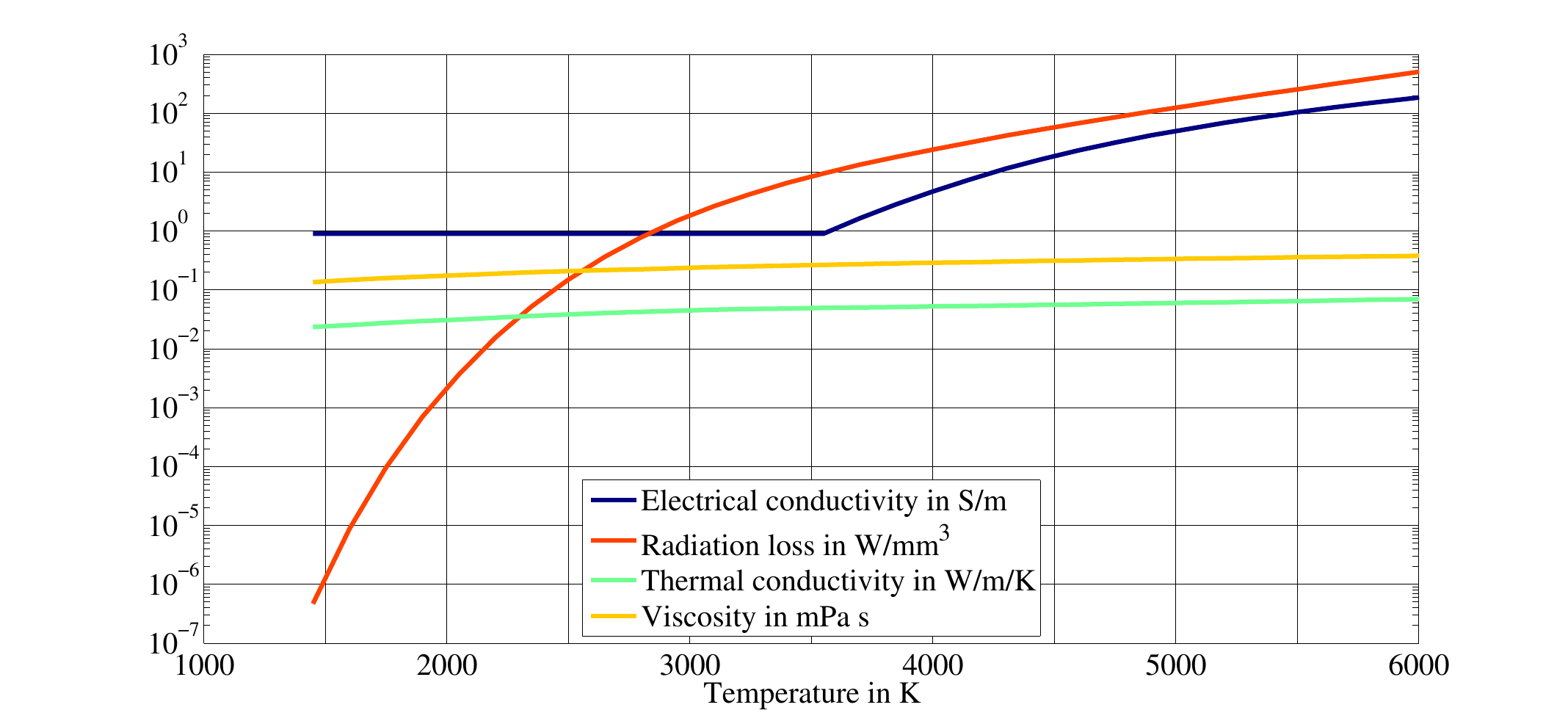}
\figcapt{}{Temperature dependency of some material properties of the arc tube filling. To account for deviations from local thermal equilibrium of the electric conductivity, a temperature-independent electric conductivity is used below a temperature threshold of 3550 K \cite{schwiegerinfluence}.
\label{fig:Materialdaten1}}
\end{figure}

\begin{figure}[h]
\centering
\includegraphics[width=0.5\linewidth,trim=0 0 0 0,clip]{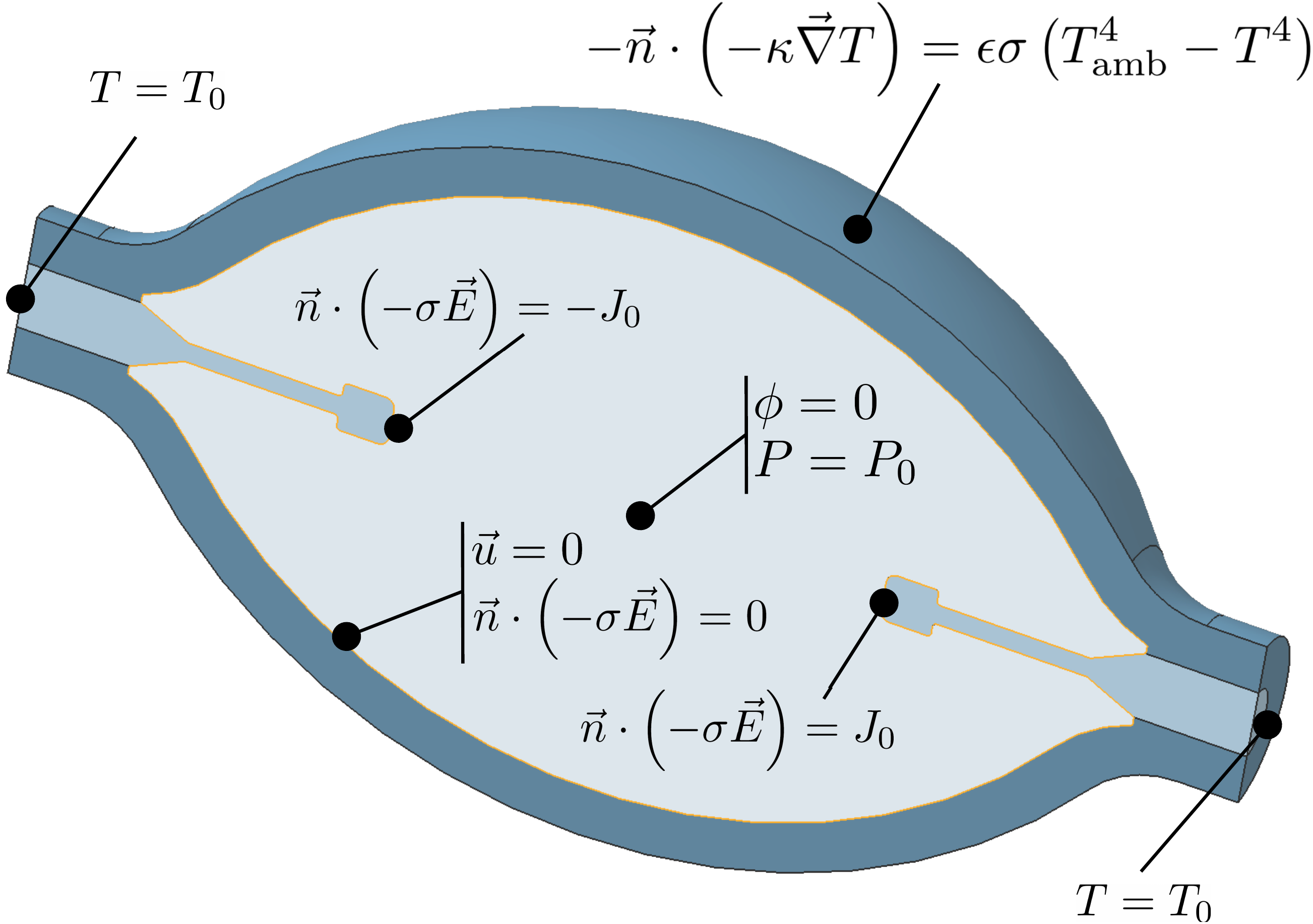}
\figcapt{}{Boundary conditions for differential equations of section \ref{sec:TemperatureField1}. \label{fig:BC1}}
%\caption{Boundary conditions for differential equations of section \ref{sec:TemperatureField1}}
\end{figure}

The electric potential $\phi$ is determined from the equation of charge conservation
\fm{\vec{\nabla}\cdot\left(\sigma\vec{E}\right)=0.}
$\vec{E}=-\vec{\nabla}\phi$ is the electric field and $\sigma$ the temperature-dependent electric conductivity\footnote{For a list of all variables see the table in the appendix.}. The temperature dependency of $\sigma$ and some other material coefficients is displayed in Figure \ref{fig:Materialdaten1}. At stable operation the velocity field $\vec{u}$ is purely buoyancy driven and can be calculated from the Navier-Stokes equation
 \mzfm{\lefteqn{\rho\left(\vec{u}\cdot\vec{\nabla}\right)\vec{u}= \vec{f}}\\ & &+\vec{\nabla}\cdot\left[-P\fif{I}+\eta\left(\vec{\nabla}\vec{u}+\left(\vec{\nabla}\vec{u}\right)^{\rm T}\right)-\frac{2}{3}\eta\left(\vec{\nabla}\cdot\vec{u}\right)\fif{I}\right]}
 in combination with the equation describing mass conservation
\fm{\vec{\nabla}\cdot(\rho\vec{u})=0}
($\rho$ density, $P$ pressure inside arc tube, $\eta$ dynamic viscosity). The buoyancy force density\footnote{Actually this term represents the  gravitional force. It is responsible for buoyancy.}
\fm{f_l=-\delta_{l3}\rho g}
($\delta_{l3}$ Kronecker symbol, $g=\ze{9.81}{m/s^2} $) bends the plasma arc upward off the symmetry axis if the lamp is operated horizontally. This loss of symmetry enforces the use of a 3D model. The temperature field $T$ inside the arc tube is determined from the Elenbaas-Heller equation
\fm{\vec{\nabla}\cdot\left(-\kappa\vec{\nabla}T\right)+\rho c_p\vec{u}\cdot\vec{\nabla}T=\sigma |\vec{E}|^2-q_{\rm rad}}
($\kappa$ thermal conductivity, $c_p$ heat capacity at constant pressure, $q_{\rm rad}$ power loss density due to radiation). The domain of the Elenbaas-Heller equation comprises the inside of the arc tube, the electrodes and the wall. The geometry of the electrodes has been modeled with slight simplifications. The differential equations have to be supplemented by boundary conditions. These can be found in Figure \ref{fig:BC1}.

The equations above serve for the calculation of the distribution of the  temperature and the power density of heat generation inside the arc tube. Both are necessary for the calculation of the acoustic pressure.

\subsection{Acoustic Pressure}
\label{sec:AcousticPressure1}
To obtain the acoustic pressure $p$ inside the arc tube, we need to solve the inhomogeneous Helmholtz equation
\fm{\vec{\nabla}\left(\frac{1}{\rho}\vec{\nabla} p\right)+ \frac{\omega^2}{\rho c^2} p
={\rm i}\omega\frac{\gamma-1}{\rho c^2}{\cal H} 
\label{eq:HelmholtzEq2}}
under the assumption that the walls of the arc tube are sound hard.
The density $\rho$ as well as the speed of sound $c$ are temperature-dependent and, therefore, space-dependent quantities\footnote{Actually  $c\rho^2$ is constant.}. 
$\gamma$ is the ratio of the heat capacities and ${\cal H}$ the power density of heat generation. In the present context we have
\fm{{\cal H}=\sigma |\vec{E}|^2-q_{\rm rad}.}

The inhomogeneous Helmholtz equation can be solved by an eigenmode expansion of the acoustic pressure \cite{kreuzer1977physics,baumann2007finite}:
\fm{p(\vec{r},\omega)=\sum_{j}^{}{A_j(\omega)p_j(\vec{r})}.\label{eq:solution1}}
The eigenmodes are obtained by solving the homogeneous Helmholtz equation and normalize the solutions according to
\fm{\int_{V_{\rm C}}{}{p_i^\ast p_j}{{\;\rm d}V}=V_{\rm C}\delta_{ij}\label{eq:normalization1}} 
($V_{\rm C}$ is the volume enclosed by the walls of the arc tube).
In this article the pressure at a certain resonance frequency $\omega_j$ is of interest. Under the assumption that the eigenfrequencies are well separated from one another the series above reduces to one term:
\fm{p(\vec{r},\omega_j)\approx{A_j(\omega_j)p_j(\vec{r})}.\label{eq:solution2}}
The amplitude can be calculated from
\fm{A_j(\omega_j)=
\frac{(\gamma-1)}{\omega_j L_j V_{\rm C}} \int_{V_{\rm C}}{}{p_j^\ast {\cal H}}{{\;\rm d}V},\label{eq:excAmplitude1}}
where $L_j$ is the loss factor. How to estimate the loss factor, is described elsewhere \cite{baumann2009finite}. Here, we consider volume loss due to heat conduction and viscosity as well as surface loss due to heat conduction and viscosity.

\subsection{Acoustic Streaming Field}
\label{sec:StreamingField1}
Fluid streaming around a rigid structure can generate noise. Less known is the opposite effect: Noise can produce fluid flow with a non-vanishing time average of mass transport. This nonlinear second order effect is called acoustic streaming \cite{Rayleigh.1883}.

If a standing pressure wave is excited in a closed vessel like the arc tube of an HID lamp, the particles of the fluid experience a viscous force, particularly those near the wall. The particles in immediate neighborhood of the wall are at rest and cannot participate in the oscillation (no-slip boundary condition). The viscosity induces a vortex-like motion of the arc tube filling inside the viscous boundary layer (inner streaming)  \cite{boluriaan2003acoustic}. Simultaneously, a second vortex-like motion is generated outside the boundary layer (outer streaming or Rayleigh streaming) \cite{boluriaan2003acoustic}. The size of the outer streaming vortices is of the order of the wavelength of the standing pressure wave. Arc flicker is caused by the outer streaming vortices \cite{afshar2008theory}. 

To obtain the streaming field, it is necessary to solve the Navier-Stokes equation
 %\mzfm{\lefteqn{\rho\left(\vec{v}\cdot\vec{\nabla}\right)\vec{v}=\vec{f}}\\ & &+\vec{\nabla}\cdot\left[-P\fif{I}+\eta\left(\vec{\nabla}\vec{v}+\left(\vec{\nabla}\vec{v}\right)^{\rm T}\right)-\frac{2}{3}\eta\left(\vec{\nabla}\cdot\vec{v}\right)\fif{I}\right]}
with force density
\fm{f_l=\frac{\partial\overline{\rho v_kv_l}}{\partial x_k}-\delta_{l3}\rho g.}
Here, Einstein's sum convention and time averaging over one cycle has to be applied. $\vec{v}$ is the sound particle velocity. For time harmonic waves the force density can be expressed by the amplitude $\vec{\hat{v}}$ of the  sound particle velocity
\fm{f_l=\frac{1}{2}\frac{\partial\rho \hat{v}_k\hat{v}_l}{\partial x_k}-\delta_{l3}\rho g,}
which can be calculated from the acoustic pressure \cite{temkin1981elements}:
\fm{\vec{\hat{v}}(\vec{r},\omega_j)=\frac{1}{{\rm i}\omega_j\rho}\vec{\nabla}p(\vec{r},\omega_j)=\frac{A_j(\omega_j)}{{\rm i}\omega_j\rho}\vec{\nabla}p_j(\vec{r}).}
Using the acoustic pressure modes from Section \ref{sec:AcousticPressure1} for the calculation of the sound particle velocity is not sufficient for the determination of the streaming field. The reason is that the eigenmodes of the Helmholtz equation do not account for the no-slip boundary condition, which is essential for the formation of the streaming vortices. The solution to this problem is to multiply the sound particle velocity by a factor that is equal to one almost everywhere inside the arc tube and drops to zero at the wall \cite{Schuster.1940}:
\fm{\hat{v}_k(\vec{r},\omega_j)\rightarrow \hat{v}^\ast_k(\vec{r},\omega_j):=\hat{v}_k(\vec{r},\omega_j)h(d).}
Here $d$ is the perpendicular distance of the point $\vec{r}$ to the wall. The function $h(d)$ is defined by
\fm{h(d)=1-\exp\left(-(1+{\rm i})d/\delta\right),}
where the viscous penetration depth is
\fm{\delta=\sqrt{\frac{2\eta}{\rho\omega_j}}.}
The function $h(d)$  contains an oscillating and a damping factor. Instead of $\hat{v}_k$, the modified sound particle velocity $\hat{v}^\ast_k$ has to be used in the force term $f_l$ of the Navier-Stokes equation.

\section{Implementation}

\begin{figure}[th]
\centering
\includegraphics[width=0.5\linewidth,trim=0 0 0 0,clip]{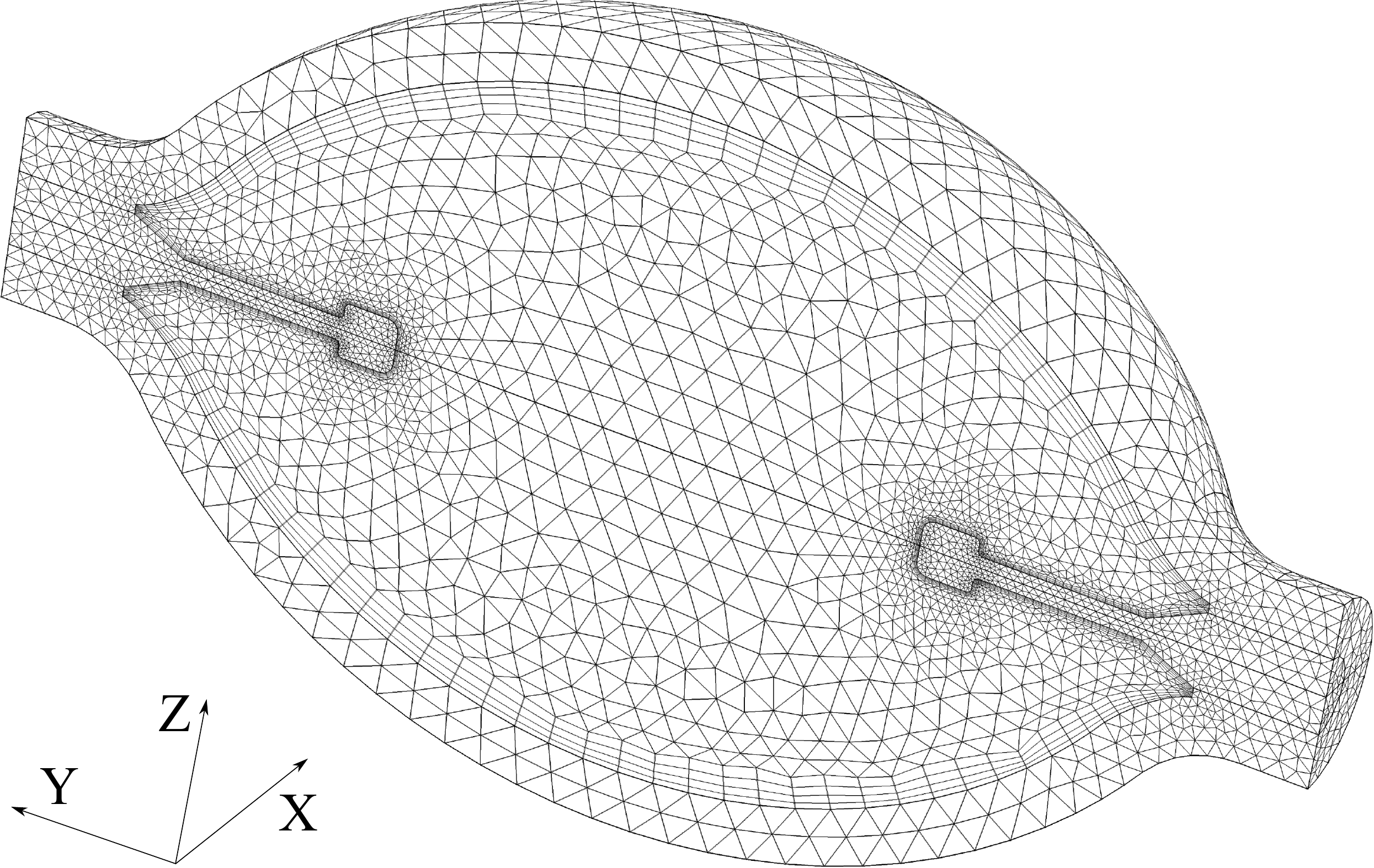}
\figcapt{}{Finite element mesh with boundary layer used for the calculation of the AS velocity field. The coordinate system used in this article has its origin centrally between the electrodes and the axes are directed as indicated. The gravitational force points to the negative $z$-direction.
\label{fig:FE-Mesh1}}
%\caption{Boundary conditions for differential equations of section \ref{sec:TemperatureField1}}
\end{figure}

The stationary temperature field that has been calculated as described in section \ref{sec:TemperatureField1} is symmetrical with respect to the vertical $y$-$z$-plane (Figure \ref{fig:FE-Mesh1}). Therefore, it is sufficient to consider one half of the physical geometry (Figure \ref{fig:BC1}) subject to appropriate symmetry boundary conditions ($\vec{n}\cdot\vec{u}=0$, $\vec{n}\cdot(-\kappa\vec{\nabla}T)=0$). In principle, the temperature field resulting from AC operation is required. In order to save computing time, corresponding stationary simulations have been performed for direct current (DC) operation. {At the anode a positive and at the cathode a negative current density with the same magnitude has been applied to preserve the model symmetry. The electric ground has been defined at the center point of the model.}
% This introduces a certain asymmetry with respect to the $x$-$z$-plane (see Figure \ref{fig:BC1}).To compensate this asymmetry, the average of the DC-field and its mirrored field has been calculated.
The fields obtained for this simplified model have been mapped onto the full geometry. For the calculation of the acoustic response function and the streaming field the full geometry has been used. Exemplary, the mesh for the calculation of the streaming field is displayed in Figure \ref{fig:FE-Mesh1}.

The simulation of the stationary step was tested with seven different finite element mesh resolutions to find the best compromise between accuracy of the results and computing time. The results (temperature, electric potential, fluid velocity) obtained with the mesh selected for the simulations presented in this article differ from the results obtained with the finest mesh by less than $1\%$. By using the coarser mesh instead of the finest mesh the computing time can be reduced by a factor of $10$.

It has been a certain challenge to obtain the distance $d$ mentioned at the end of Section \ref{sec:StreamingField1}. This was achieved by defining a subdomain of thickness $\ze{50}{\mu m}\approx 10\delta$ inside the arc tube and adjacent to the wall. For this subdomain a heat transfer model for solids has been introduced. At the boundary to the wall a zero temperature boundary condition has been implemented. At the opposite boundary the temperature has been set to one. The pseudo temperature resulting from this heat transfer problem varies linearly from zero to one across the sub-domain  and can, after scaling with the proper factor, be used as an estimate for $d$. The mesh of Figure \ref{fig:FE-Mesh1} is too coarse to resolve the function $h(d)$ properly, in particular to resolve the oscillating factor. The important property of this function in the present context is the exponential descent towards the wall of the arc tube. The mesh is fine enough to resolve this descent, albeit only coarsely.

\section{Results}

\subsection{Acoustic Streaming Field}
\label{sec:ResultsStreamingFiled}
%In Figure \ref{fig:3D_Temperature1} the temperature distribution is depicted. This temperature field serves as input for the calculation of the acoustic eigenmode (Figure \ref{fig:47k1AcPressureMode1}).
%\begin{figure}[htb]
%\centering
%\includegraphics[width=0.90\linewidth,trim=0 0 10 0,clip]{3D_TemperatureT2+Skala} 
%\figcapt{}{Temperature field. The numbers at the color scale indicate the temperature in $\rm K$.
%\label{fig:3D_Temperature1}}
%\end{figure}

\begin{figure}[bth]
\centering
\includegraphics[width=0.50\linewidth,trim=0 0 0 0,clip]{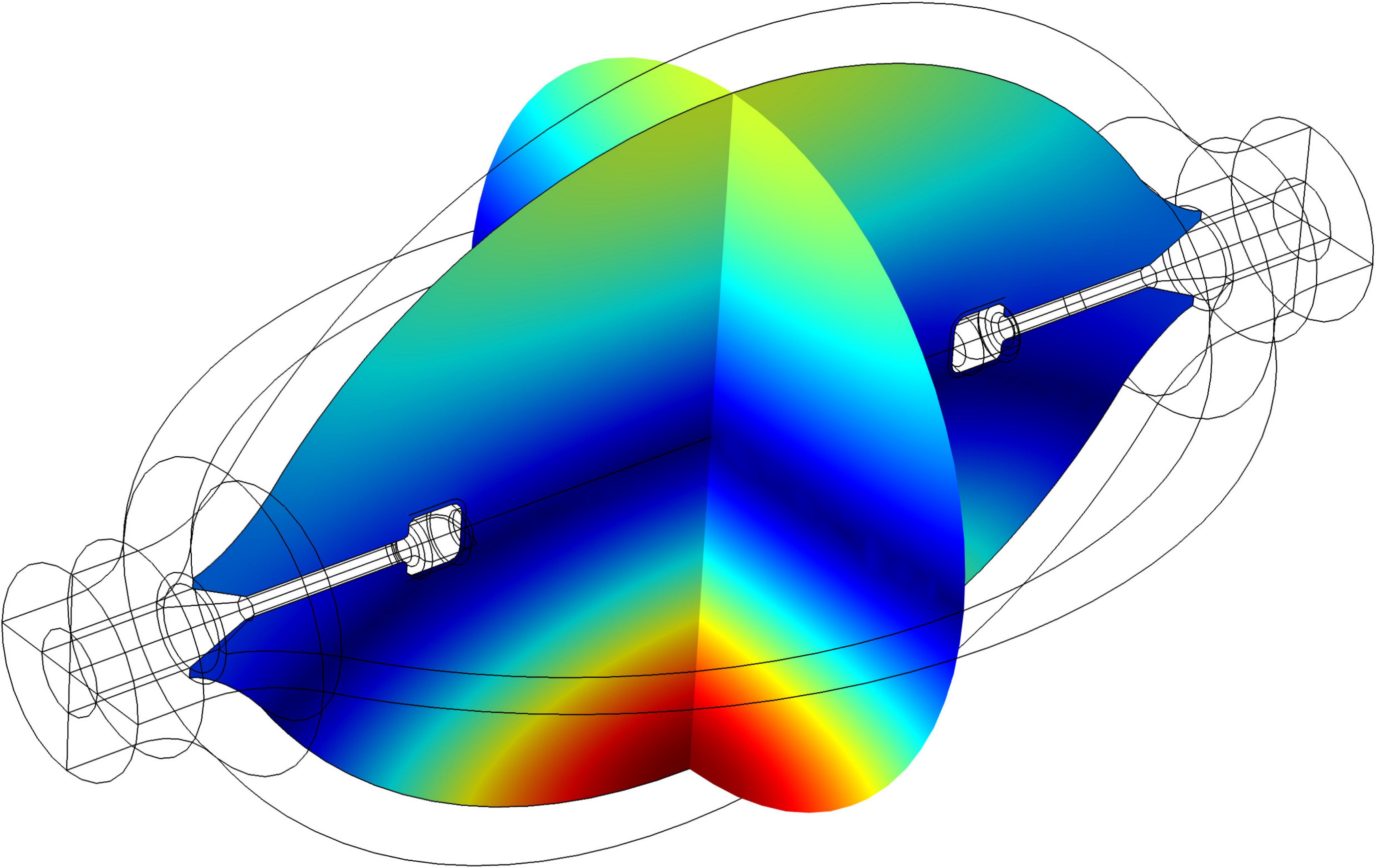} 
\figcapt{}{Absolute value of the acoustic pressure for the $\ze{47.4}{kHz}$ mode. Blue indicates $|p|=0$ and red the maximal value of $|p|$. The actual values are of no importance.
\label{fig:47k1AcPressureMode1}}
\end{figure}

\begin{figure}[th]
\centering
\includegraphics[width=0.55\linewidth,trim=0 0 0 0,clip]{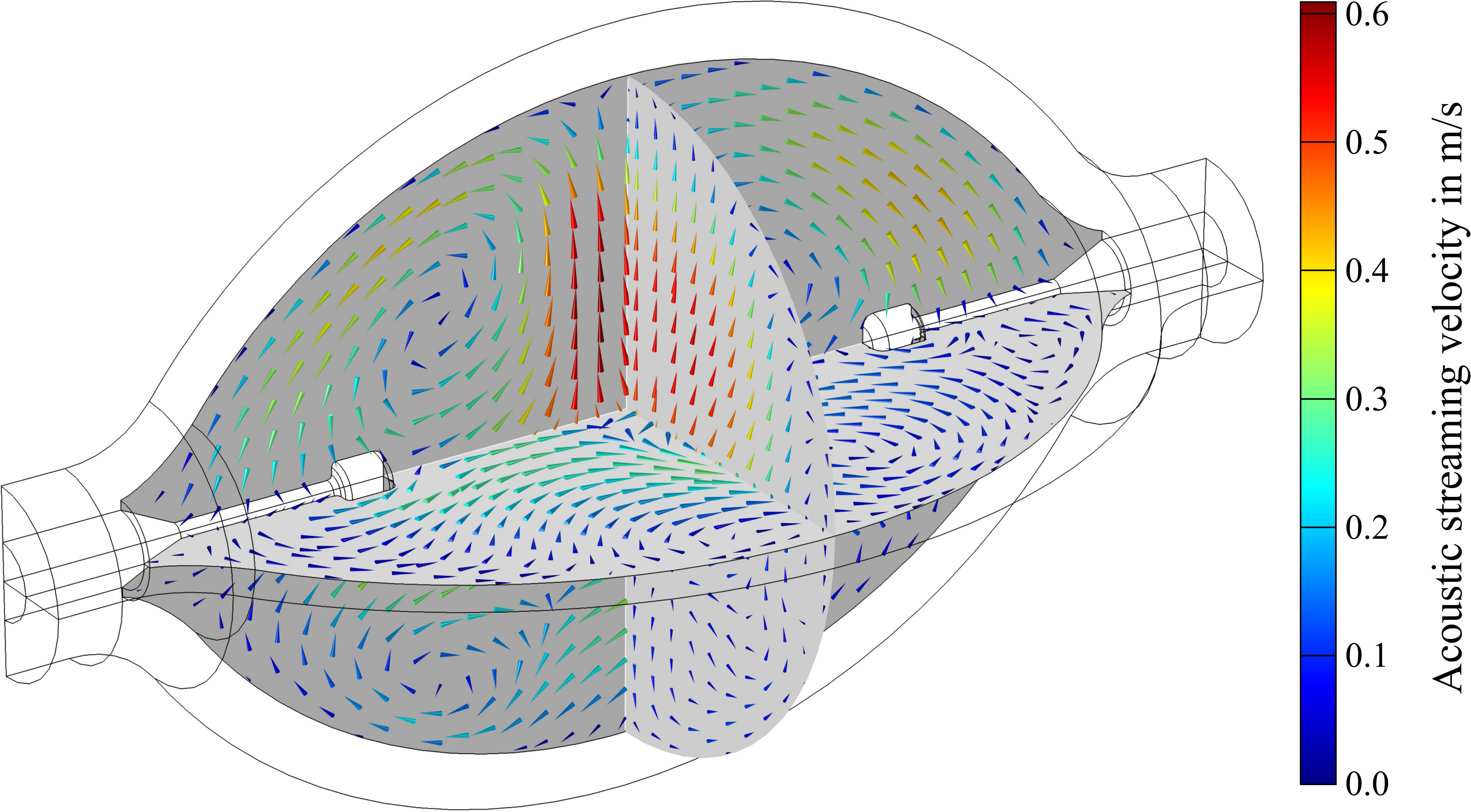} \\
\vspace{10mm}
\includegraphics[width=0.45\linewidth,trim=0 0 0 0,clip]{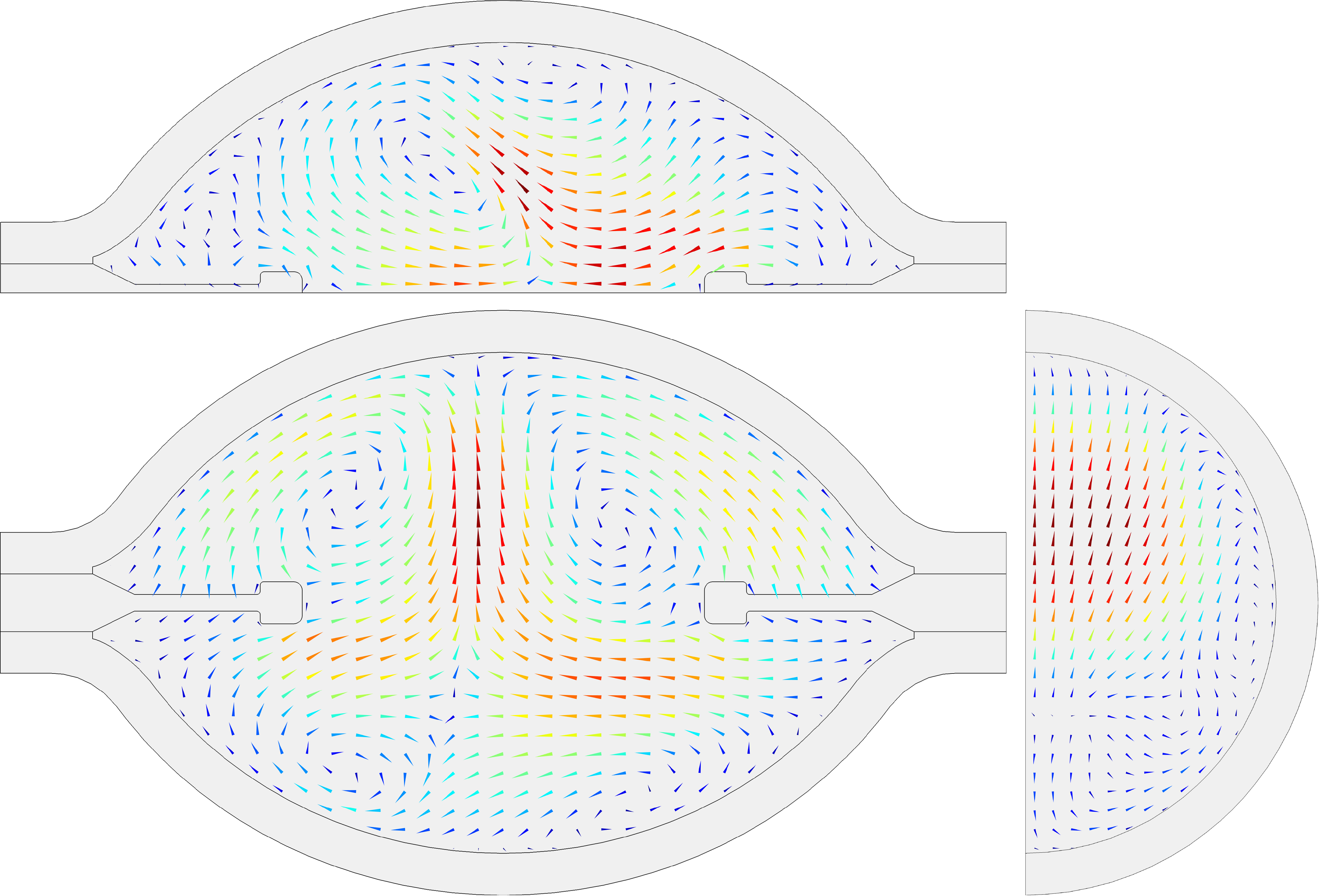}
\figcapt{}{Three dimensional illustration of the AS velocity field (top)  and the three planes separately (top: $x$-$y$-plane; bottom left: $y$-$z$-plane; bottom right: $x$-$z$-plane).\label{fig:AcousStreamingVelocity1}}
\end{figure}

Figure \ref{fig:47k1AcPressureMode1} shows the profile of the acoustic pressure mode we investigated in this paper. In Figure \ref{fig:AcousStreamingVelocity1} the AS field associated with this mode is depicted. {It has been obtained by the procedure described in Section \ref{sec:SymmetryBreaking1}.} Figure \ref{fig:KundtTube1} shows the AS flow pattern due to a longitudinal acoustic mode in a closed cylindrical tube with a length equal to half a wavelength. This velocity field has been derived analytically about 75 years ago \cite{Schuster.1940}. A close look at the flow pattern in the $y$-$z$-plane of Figure \ref{fig:AcousStreamingVelocity1} reveals the same structure as the one in Figure \ref{fig:KundtTube1} - namely two  vortices in the upper part and two vortices in the lower part (upper right and bottom left: clockwise; bottom right and upper left: counter-clockwise). The two systems are alike in principle but different in details like the shape of the vessel containing the fluid, the temperature and  buoyancy force distribution in space as well as the pressure distribution of the acoustic mode. That we observe a similar structure of the flow patterns, indicates that the finite element model works properly.

%\begin{figure}[th]
%\centering
%\includegraphics[width=0.99\linewidth,trim=0 0 6 0,clip]{ASVelCompBuoy_100pc_3D+Skala} 
%\figcapt{}{Streaming velocity field. The numbers at the color scale indicate the velocity in $\rm m/s$.
%\label{fig:3D-AcousStreamingVelocity}}
%\end{figure}

Figure \ref{fig:AcousStreamingVelocity1} shows a maximal AS velocity of approximately {$\ze{0.61}{m/s}$}. Dreeben reported  a maximal velocity of $\ze{1}{m/s}$ from his 2D model \cite{dreeben2008modelling}. The maximal buoyancy velocity in the 3D model is {$\ze{8.5}{cm/s}$} compared to $\ze{8}{cm/s}$ in the 2D model. Since the investigated lamps are different with respect to geometry, wattage and other properties, this is a very satisfying accordance.

The AS field dominates over the buoyancy driven flow. It is clear that the physical conditions inside the arc tube change drastically once AS sets in. Under the influence of the altered flow field the arc moves to a new position and the temperature distribution changes substantially from the original field. Therefore, the space-dependent speed of sound and other physical properties also change. As a result the frequency of the eigenmode is changed and shifted away from the AC power frequency. The AS force diminishes or even vanishes. Temperature field, speed of sound and arc resume their original position and the process starts all over again, i.e. the arc flickers \cite{afshar2008theory}. 

At present the feedback of AS on the temperature field is not included in the model, but it can be used to investigate the influence of the AS field and, therefore, light flicker, under certain operation conditions. In a next step the calculations have to be repeated recursively to obtain the correct stationary fields. The AS force has to be calculated from the sound particle velocity of the previous step. Once convergence has been obtained, it would be interesting to investigate, whether the resulting flow field is stable or not. This can be accomplished with the aid of a linear stability analysis \cite{cross2009pattern}. It seems natural to link the transition to instability with the onset of flicker. To solve the flicker problem directly with a transient FE model, is by orders of magnitude more demanding regarding computational requirements.

\subsection{Symmetry Breaking and Critical Behavior}
\label{sec:SymmetryBreaking1}
The flow pattern depicted in Figure \ref{fig:AcousStreamingVelocity1} is not mirror symmetric with respect to the $x$-$z$-plane (see Figure \ref{fig:FE-Mesh1}). The model itself is symmetric and the question arises, how this asymmetry can be explained.

It is well known that in nonlinear dynamical systems, like the one under investigation here, {far from thermodynamic equilibrium} symmetry breaking can occur \cite{cross2009pattern}. {In these dissipative systems entropy is transfered to the surroundings and decreases locally.} Famous examples are the Taylor vortices, which can appear in the gap between rotating cylinders (Taylor-Couette system), and the Rayleigh-B\'{e}nard system. In HID lamps bifurcation points of the current transfer to arc cathodes have been investigated \cite{Benilov1998,BenilovCunha2003}. In these systems a symmetry gets lost once a certain control parameter exceeds a critical value.

In the Taylor-Couette and the Rayleigh-B\'{e}nard system the loss of symmetry is due to the exertion of opposing forces on the fluid elements. Taylor vortices form, once the centrifugal force prevails over the viscous force. In the Rayleigh-B\'{e}nard system certain structures emerge once thermal diffusion and viscous force are not able to balance the buoyancy force anymore. In both cases a highly symmetrical state of the fluid becomes unstable and a phase transition of the second kind to a new state of less symmetry occurs. The new state is characterized by certain patterns depending on the system under consideration.

\begin{figure}[th]
\centering
\includegraphics[width=0.99\linewidth,trim=0 0 0 0,clip]{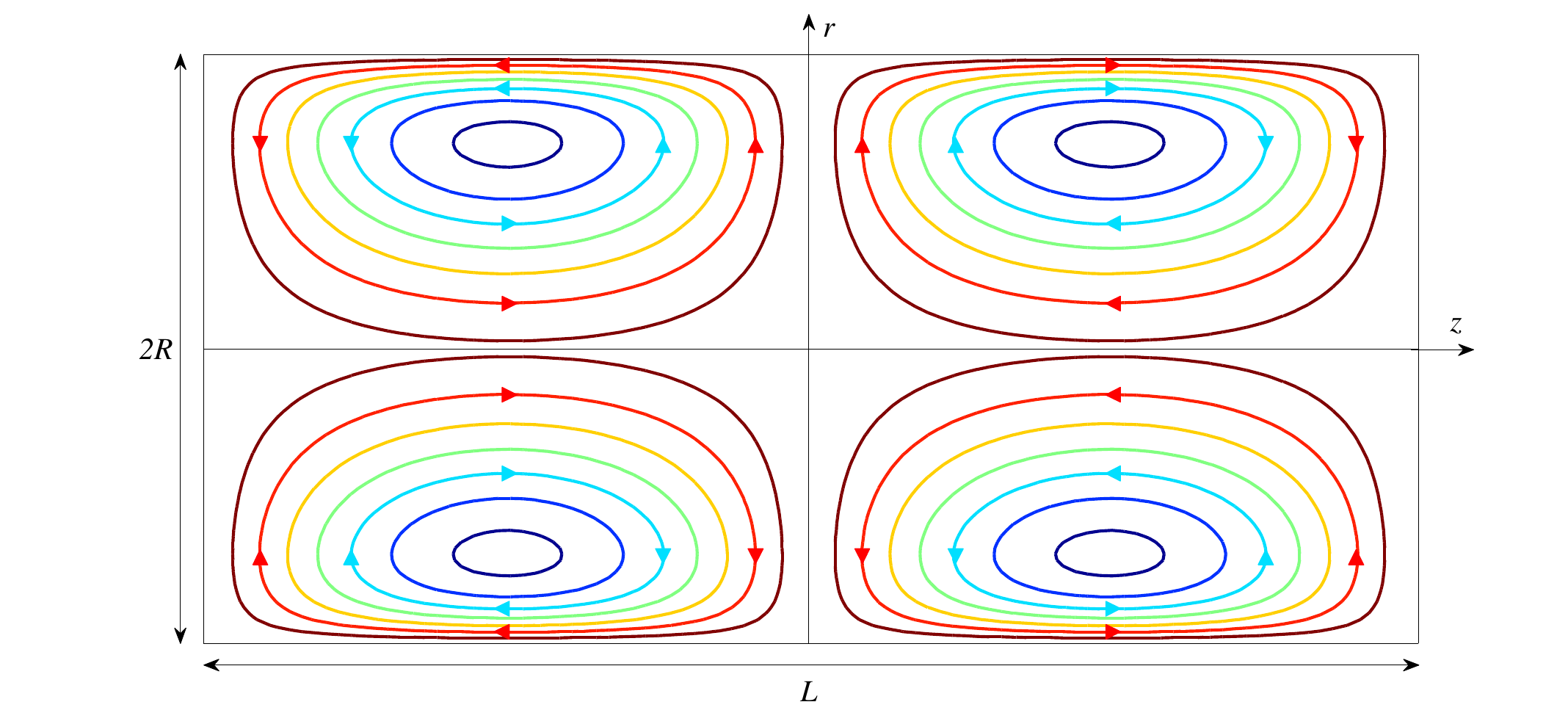}
\figcapt{}{AS velocity field resulting from longitudinal modes in a cylinder \cite{Schuster.1940}. $r$ and $z$ denote cylinder coordinates in arbitrary units.
\label{fig:KundtTube1}}
\end{figure}

If standing waves are excited in a cylinder, the AS velocity field depicted in Figure \ref{fig:KundtTube1} develops. Once the acoustic pressure amplitude exceeds a critical value, the AS force is larger than the viscous force \cite{Schuster.1940}. Before AS sets in, the fluid field $\vec{u}$ in the cylinder is identical to zero (continuous translational symmetry). Above the critical value a pattern of vortices is present (in a cylinder of infinite length this corresponds to a discrete translational symmetry).

In case of the HID lamp we observe a similar situation. Assume for the moment that the lamp is operated in outer space, where there is no gravity and the buoyancy force is zero. Consequently, the fluid velocity $\vec{u}$ is zero inside the arc tube. Near an acoustic resonance the AS force might become large in comparison to the viscous force. Except for the shape of the vessel and the temperature variation inside the arc tube, the situation is similar to the standing waves in a cylinder. Thus, an instability from a translational symmetric state ($\vec{u}=\vec{0}$) to a new state, which shows a pattern of vortices, would be observed in outer space due to the interaction of the AS and the viscous forces. The detailed appearance of the vortex pattern depends on the acoustic mode under consideration.

In the following it is assumed that the lamp is operated horizontally under the influence of gravity. We calculated a series of AS fields with the force term
\fm{f_l=S\frac{\partial\overline{\rho v_kv_l}}{\partial x_k}-\delta_{l3}\rho g,}
where $0<S\le1$ is used as a control parameter. This could be realized experimentally by shifting the AC frequency away from
the resonance frequency or lowering the modulation depth \cite{hirsch2010acoustic}. This should have the same effect, namely the reduction of the AS force. For $S=0$ (no AS)  the fluid elements in the center of the arc tube experience the buoyancy force in upward direction, which results in the formation of two vortices. 

\begin{figure}[t]
\centering
\includegraphics[width=0.99\linewidth,trim=50 0 0 0,clip]{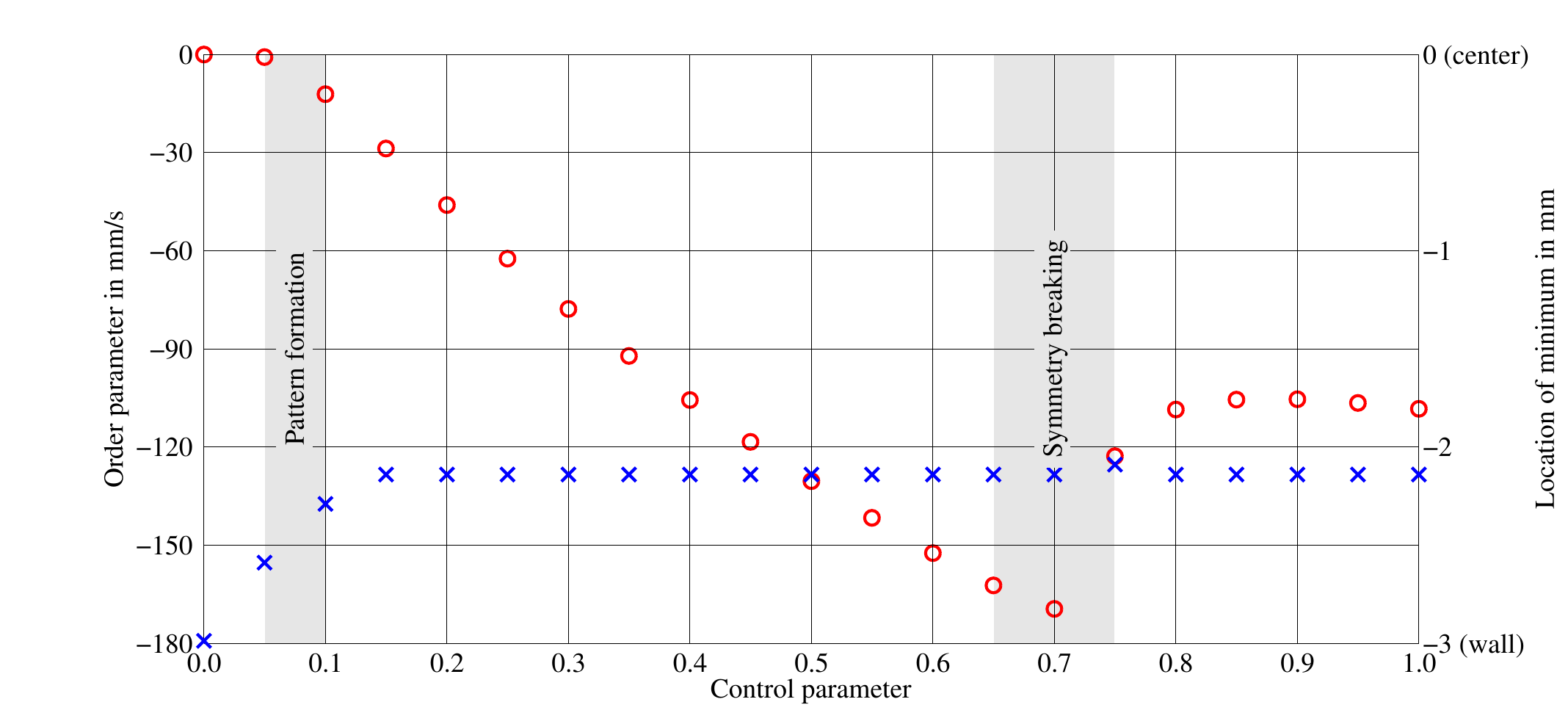}
\figcapt{}{Order parameter $\Psi$ (circles, left axis) and location of  $u_z(0,0,z)$ minimum (crosses, right axis) vs. control parameter $S$. When increasing the control parameter $S$ from $0$ to $1$, the formation of additional vortices sets in slightly {below $S=0.1$} (somewhere inside the shaded region labeled 'Pattern formation'). Inside the shaded region labeled 'Symmetry breaking'  the velocity field undergoes a transition from a mirror symmetric state to an asymmetrical state. As to be expected, $\Psi=0$ corresponds to the no-slip condition at the arc tube's wall.
\label{fig:OrderParameterPsi1}}
\end{figure}

For $S>0$ the AS force in the centrical upper part of the arc tube also points in upward direction\footnote{The viscous force is always opposed to the fluid velocity.}. In the center of the lower part, however, the AS force tends to move the fluid elements downward, i.e. buoyancy and AS force point in opposite directions. Once the AS force is strong enough, two additional vortices appear. The formation of these additional vortices requires that the vertical velocity component of the fluid velocity is negative on parts of the $z$-axis. Therefore, we use 
\fm{\Psi:=\min\limits_{z\in\mathcal{D}} u_z(0,0,z)}
as an order parameter for this type of pattern formation. $\mathcal{D}$ is the part of the $z$-axis inside the arc tube. In Figure \ref{fig:OrderParameterPsi1} the order parameter $\Psi$ is depicted as function of the control parameter $S$ together with the value of $z$ where the minimum of $u_z$ occurs. As expected from the reasoning above, the velocity $u_z$ is non negative at low values of $S$ (dominance of buoyancy) and becomes negative when $S$ increases (dominance of AS). A negative value of $\Psi$ indicates that the original state has become unstable and two additional vortices have been created. The value of the control parameter, at which the order parameter $\Psi$ becomes negative, is called critical point. From the figure we conclude that the critical point for this type of instability is near {$0.05$\footnote{Technically speaking it is not clear, if the term \emph{critical point} is justified. A critical point is characterized by a singularity of the order parameter and we do not observe any signs of a singularity for $S\approx 0.05$.}}.

% In Figure \ref{fig:vertASVelo1} the vertical component of the fluid velocity $u_z$ at various points in the center of the lower part of the arc tube as a function of the control parameter $S$ is displayed. As expected from the reasoning above,  the velocity $u_z$ is non negative at low values of $S$ (dominance of buoyancy) and becomes negative when $S$ increases (dominance of streaming). Once $u_z$ becomes negative two additional vortices are forming (pattern formation).

Figure \ref{fig:OrderParameterPsi1} reveals that the flow behavior changes at {$S\approx 0.7$}. In the following we show that the {jump} of $\Psi(S)$ is related to the loss of the mirror symmetry. To quantify the  loss of symmetry, a second order parameter is introduced. We choose
\fm{\Phi:=\frac{1}{V_{\rm r}}\int_{V_{\rm r}}{}{\left|u-\tilde{u}\right|}{{\;\rm d}V,}\label{eq:orderParam1}} 
in which the integral is over the right half of the interior of the arc tube. $u(x,y,z)$ is the absolute value of the fluid velocity and $\tilde{u}(x,y,z)$ is $u(x,-y,z)$. $\Phi$ measures the averaged difference of the velocities at the opposing points $(x,y,z)$ and $(x,-y,z)$ inside the arc tube. Therefore, the value of $\Phi$ is a measure for the asymmetry. In the symmetric phase $\Phi$ is zero. In the case of broken symmetry $\Phi$ assumes a positive value. The value of the control parameter, at which symmetry breaking sets in, is a second critical point.

\begin{figure}[t]
\centering
\includegraphics[width=0.99\linewidth,trim=100 0 100 0,clip]{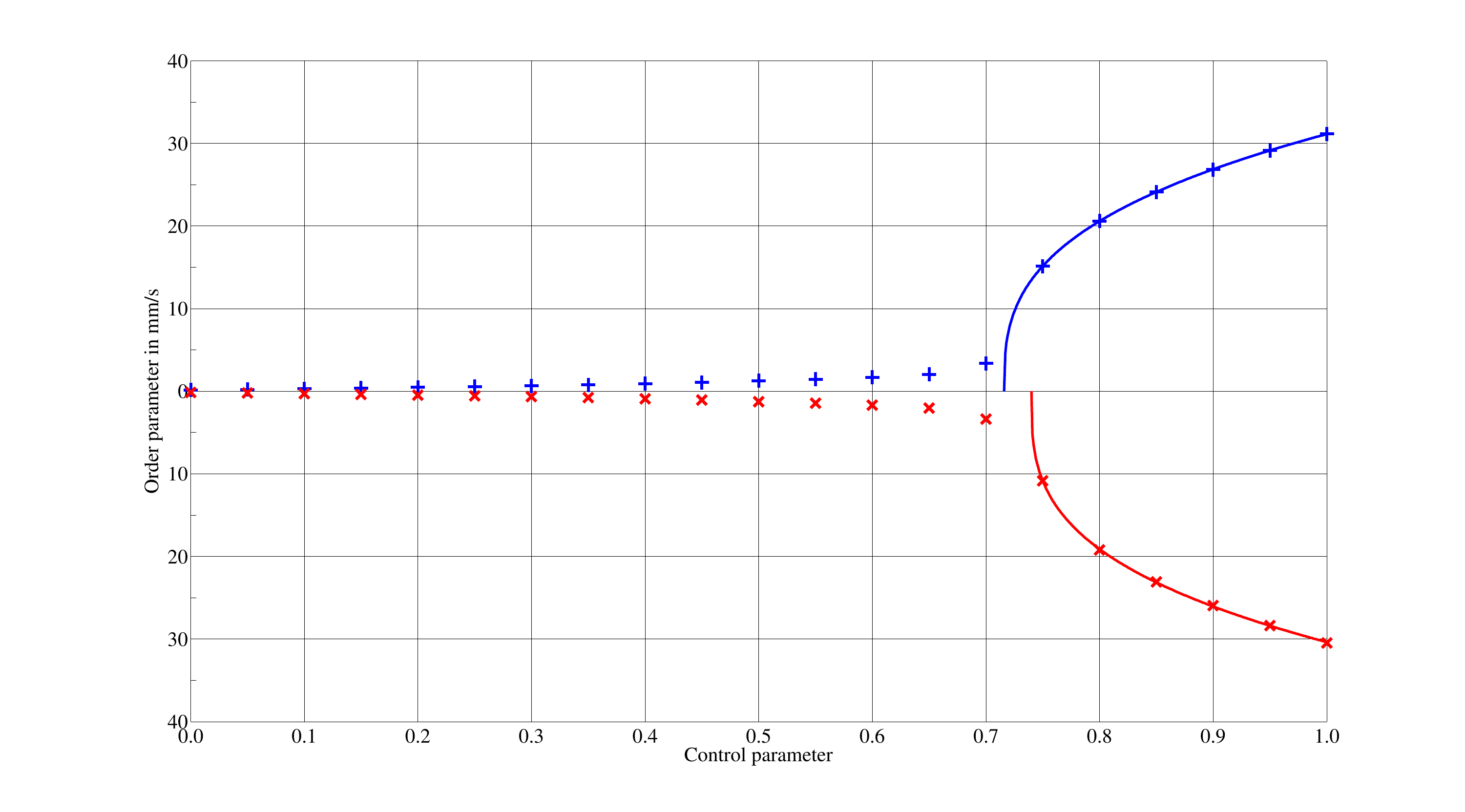}
\figcapt{}{{Order parameter $\Phi$ and fit function vs. control parameter $S$. The results in the upper part are for the approach from positive tilt angles, while the results in the lower part are for the approach from negative tilt angles. The curves have been obtained from a fit to the function $a(S-S_{\rm crit})^\beta$. At $S=1$ the averaged difference of the fluid velocity at a point $(x,y,z)$ and its mirrored point $(x,-y,z)$ is ca. $\ze{30}{mm/s}$. The deviation of $\Phi$ from zero below $S_{\rm crit}$ is assumed to be due to numerical noise. The order parameter $\Phi$ is by definition non-negative and, therefore, cancellations are not possible.}
\label{fig:OrderParameter1}}
\end{figure}

\begin{figure}[t]
\centering
\includegraphics[angle=0,width=0.55\linewidth,trim=0 0 0 0,clip]{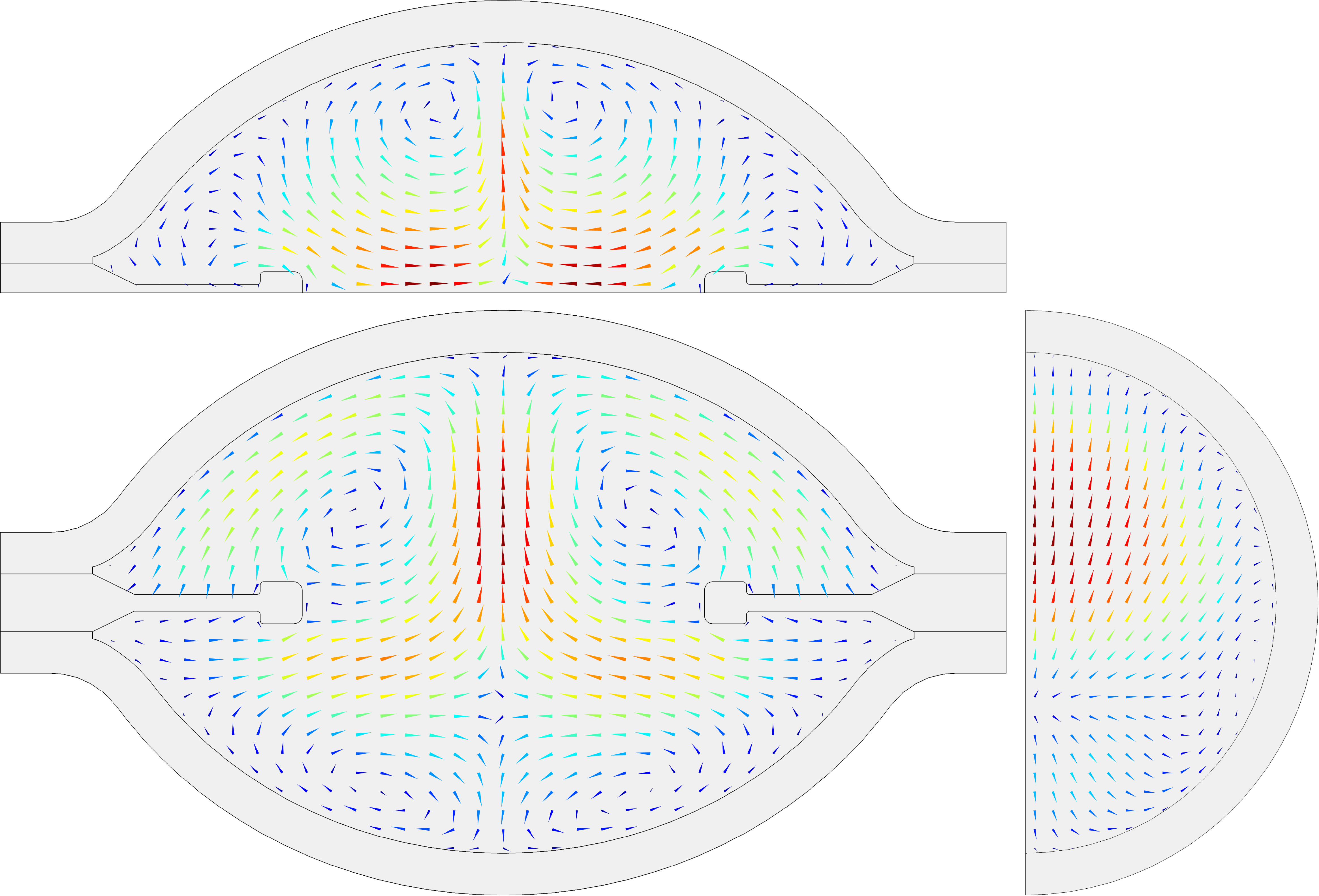}
\figcapt{}{AS velocity field obtained with downscaled AS force term ({$S=0.7$}). In this case the maximal AS force exceeds the maximal buoyancy force by a factor of approximately {$6.6$}. The figures on the left show an almost perfect symmetry.
\label{fig:AcousStreamingVelocity2}}
\end{figure}

{The order parameter $\Phi$ has been calculated as a function of $S$. For $S> 0.7$ the function $\Phi(S)$ manifests erratic fluctuations, indicating that the system has become unstable. In order to search for stable solutions, we calculated the flow field for a slightly tilted lamp (tilt angles $\alpha=\pm 10\grad$, $\pm 5\grad$, $\pm 4\grad$, $\pm 3\grad$, $\pm 2\grad$ and $\pm 1\grad$). The asymmetrical solutions resulting from the tilting have been used as initial conditions for the calculation of the AS flow field of the lamp operated at the next lower tilt angle. For the horizontally operated lamp ($\alpha=0$) the asymmetry in the initial condition leads to the stable asymmetrical solution depicted in Figure \ref{fig:AcousStreamingVelocity1}. The order parameters $\Phi$ obtained by approaching the horizontal lamp position from positive and from negative tilt angles are depicted in Figure \ref{fig:OrderParameter1}. For $S\le0.7$ the values of $\Phi$ are small and the solutions are symmetric (see Figure \ref{fig:AcousStreamingVelocity2}). Figure \ref{fig:OrderParameter1} clearly shows that the symmetry gets lost at $S\approx 0.7$.}

{The data for $S>0.7$ have been fitted to $a(S-S_{\rm crit})^\beta$. $\beta$ is called critical exponent of the order parameter.
The fit of the simulation results to the power law is excellent. The upper curve yields $S_{\rm crit}=0.716$ and $\beta=0.339$, the lower curve  yields $S_{\rm crit}=0.740$ and $\beta=0.316$. The corresponding mean values are $0.728$ and $0.3275$.}

{The fit curves give the impression that the upper and the lower curve differ substantially, but the difference is due to the values at $S=0.75$ only. All other data in the upper and the lower part are not far apart from each other. An explanation for the difference at $S=0.75$ might be that the slight asymmetry in the FE mesh results in an amplification effect near the singularity.}

The transition at the critical point leads from a mirror symmetric state to a state which does not show mirror symmetry. By tilting the lamp slightly to either side of its horizontal position, it is possible to choose between two equivalent flow states (pitchfork bifurcation). The situation is quite similar to the one of a ferromagnetic material in an external magnetic field $\vec{H}$\footnote{A further system, which shows a similar behavior, is the buckling phenomenon \cite{hunt2006buckling}.}. For $S<S_{\rm crit}$ the flow field corresponds to a ferromagnet in the para\-magnetic phase ($T>T_{\rm Curie}$). The magnetization $\vec{M}$, which plays the role of the order parameter, assumes values that are dictated by the external field $\vec{H}$. In the case of the HID lamp the tilt angle $\alpha$ (or the lateral component of the buoyancy force) determines the value of the order parameter $\Phi$. $S>S_{\rm crit}$ corresponds to the ferromagnetic phase ($T<T_{\rm Curie}$). The order parameter $\Phi$ respectively the magnetization $\vec{M}$ assumes values different from zero, even for $\alpha=0$, respectively $\vec{H}=\vec{0}$.

The most prominent microscopic model for a ferromagnet is probably the Ising model \cite{ising1925beitrag}. For the Ising model in three dimensions the generally accepted value of the critical exponent $\beta$ is $0.3265$ \cite{pelissetto2002critical}. The difference of this value to our mean value of $0.3275$ is $0.31\%$. This is remarkable. However, it seems questionable if the coincidence of the two values has any significance. The two systems are quite unlike and most probably belong to different universality classes.

\begin{figure}[th]
 \centering
 \includegraphics[width=0.95\linewidth,trim=120 50 100 0,clip]{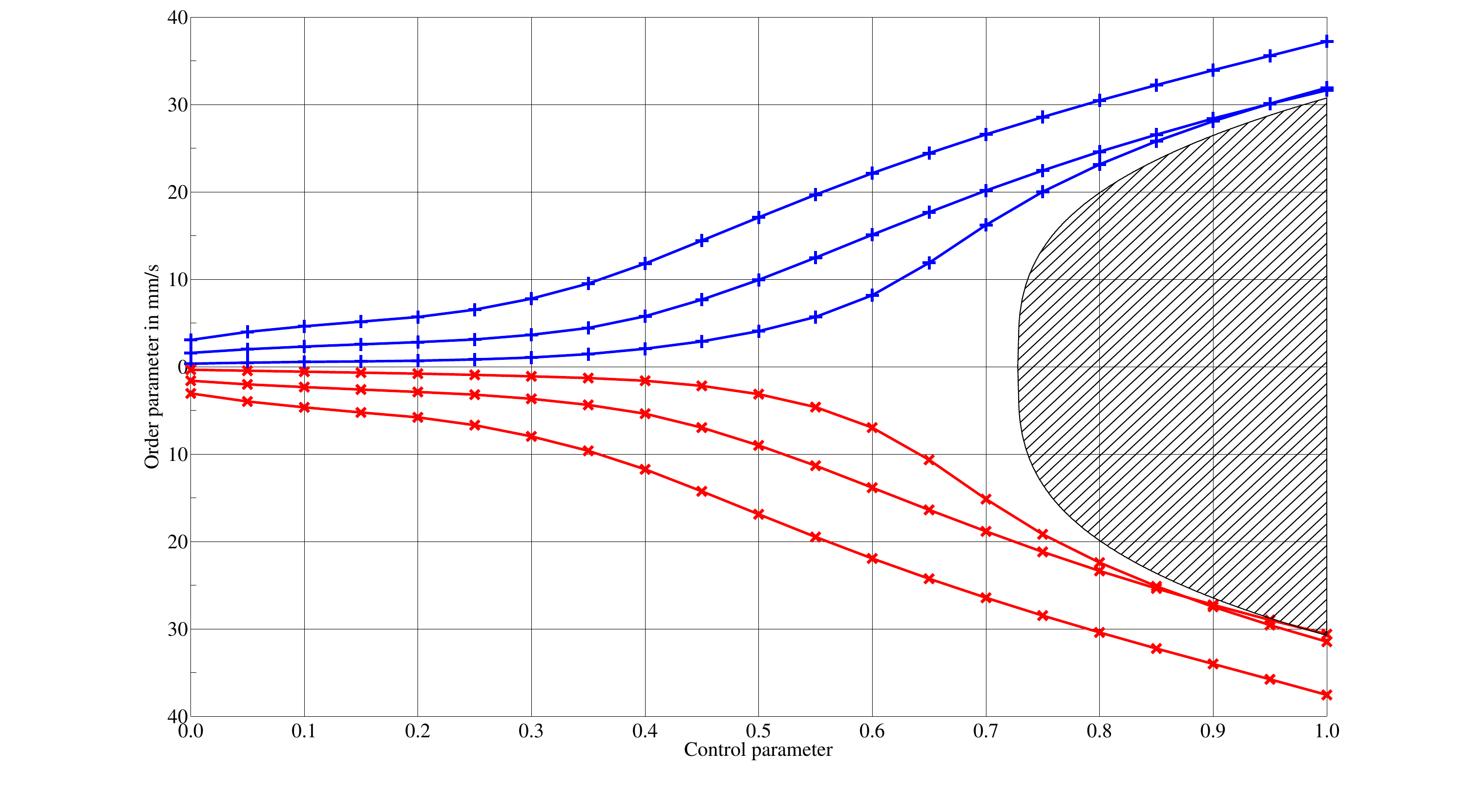}
 \figcapt{}{{Order parameter $\Phi$ vs. control parameter $S$ for the tilted lamp. The curves represent the tilt angles $\pm 10\grad$, $\pm 5\grad$ and $\pm 1\grad$. The behavior seen here is very similar to the one observed in ferromagnets when the  magnetization is depicted as a function of the temperature, with the external field $H$ as a parameter. The shaded region has been obtained from the fit function $a(S-S_{\rm crit})^\beta$ and the mean values of $a$, $\beta$ and $S_{\rm crit}$.}
 \label{fig:OrderParameter2}}
 \end{figure}

\begin{figure}[th]
 \centering
 \includegraphics[width=0.9\linewidth,trim=0 0 0 0,clip]{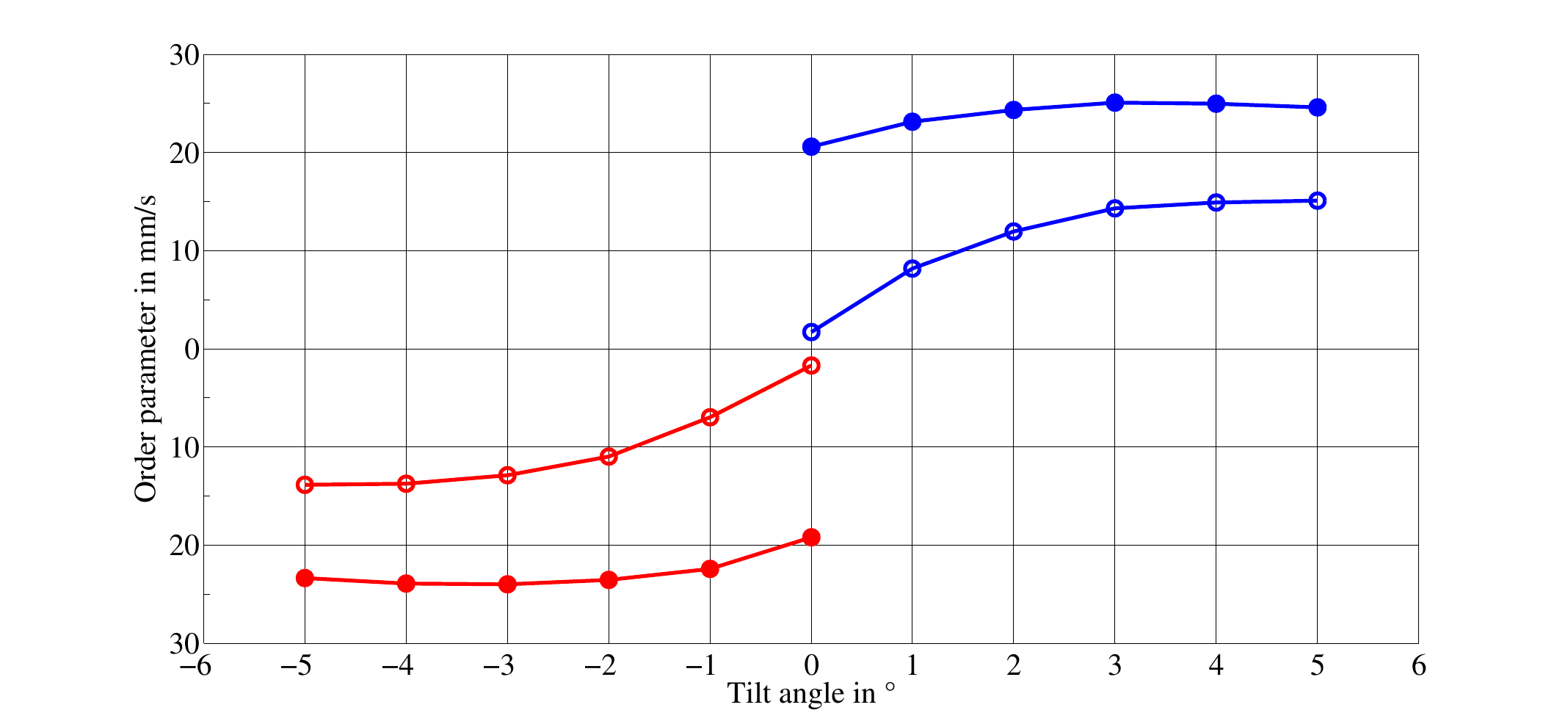}
 \figcapt{}{{Order parameter $\Phi$ vs. tilt angle $\alpha$ for $S=0.6<S_{\rm crit}$ (open circles) and for $S=0.8>S_{\rm crit}$ (full circles). Again, the jump in the $S<S_{\rm crit}$-curve is attributed to numerical noise.}
 \label{fig:OrderParameter3}}
 \end{figure}

For $\alpha\ne 0$ ($\vec{H}\ne\vec{0}$) the order parameters have contributions from symmetry breaking and from the external field respectively from lamp tilting as well. {For the HID lamp the situation is illustrated in Figure \ref{fig:OrderParameter2}. When tilting the lamp for $S>S_{\rm crit}$ from positive values of $\alpha$ to negative values or vice versa, a first order phase transition takes place (Figure \ref{fig:OrderParameter3}).}
 
\section{Conclusions}
A stationary 3D finite element model for the calculation of the AS field inside the arc tube of an HID lamp has been developed. The results obtained with the model are consistent with theoretical expectations. It has been found that the AS field suffers a symmetry breaking transition with the AS force as control parameter. 

AS is assumed to be responsible for light flicker, which is observed when the lamp is operated near an acoustic resonance frequency. The maximal AS velocity at the investigated resonance is ca. {$\ze{0.6}{m/s}$}. This is a large velocity inside a vessel with a diameter of $\ze{6}{mm}$ that corresponds to a maximal Reynolds number of approximately {$410$}. The large velocity results in an instability phase transition and symmetry breaking. It is easy to imagine that these violent changes result in arc flicker. The model can be used to investigate if light flicker at certain operation conditions can be expected. Furthermore, it enables to calculate the correct stationary fields by implementation of a recursion procedure. Finally, the flow field resulting from the recursion procedure can be examined by a linear stability analysis. We plan to extend our work in this sense in the near future.

The AS field at other acoustic eigenmodes can be used to stabilize the plasma arc (arc straightening) \cite{olsen2011experimental}. The finite element model, which is presented here, should be useful in finding these modes.
\vfill

\section{Appendix}

\renewcommand{\arraystretch}{1.5}
%\begin{table*}[tb]
\begin{tabular}{p{50mm}p{70mm}c}
\hline\noalign{\smallskip}
\noalign{\smallskip}\noalign{\smallskip}
            Electric conductivity $\sigma$ & see Figure \ref{fig:Materialdaten1} & \cite{schwiegerinfluence} \\
            Density $\rho$ & $\frac{P M}{R_{\rm m}T}$ & \\
            Static pressure $P$ &  $\ze{2.81}{ MPa}$ &\\
            Molar mass $M$ & linear function of $T$ through the points\newline $\ze{191}{ g/mol}$ at $1450\,{\rm K}$ and  $\ze{184}{g/mol}$ at $5800\,{\rm K}$ & \cite{PhilipsMaterialDatabase} \\
            Viscosity $\eta$  &  see Figure \ref{fig:Materialdaten1}  &  \cite{PhilipsMaterialDatabase} \\
            Thermal conductivity $\kappa$ of ...  &  & \\
            ... arc filling & see Figure \ref{fig:Materialdaten1} &  \cite{PhilipsMaterialDatabase} \\
             ... tungsten  & according to literature & \cite{hust1984update}  \\
             ... PCA & $(0.0378-556\cdot10^{-7}\Delta T$ ...\newline
             \hspace*{20mm}... $+282\cdot10^{-10}\Delta T^2)\cdot 10^{3}\frac{\rm W}{\rm m\,K}$ \newline with $\Delta T = T - 273.15\,{\rm K}$ & \cite{PhilipsMaterialDatabase}\\
            Specific heat capacity at\newline constant pressure $c_{p}$ &  linear function of $T$ through the points\newline $\ze{111}{\rm J/(kg\,K)}$ at $1450\,{\rm K}$ and $\ze{115}{\rm J/(kg\,K)}$\newline at $5800\,{\rm K}$ &\\
            Power loss due to radiation $q_{\rm rad}$ &  see Figure \ref{fig:Materialdaten1} & \cite{PhilipsMaterialDatabase} \\
            Temperature at electrode ground $T_0$ & $\ze{1430}{ K}$ &\\
            Emissivity of PCA & $\epsilon= \frac{(110\,t+19)\cdot 10^{12}}{T^5}+0.195-\frac{0.017}{t}$ \newline  with $t=\ze{0.5}{mm}$ (thickness of arc tube wall) &  \cite{PhilipsMaterialDatabase} \\
            Ambient temperature $T_{\rm amb}$ & $\ze{293}{K}$\\
            Current density $J_0$ & $\ze{ 0.548}{ A/mm^2}$ &\\
            Specific heat ratio $\gamma$ & $1.4$ &\\
            Speed of sound $c$ & $\sqrt{\frac{\gamma R_{\rm m}T}{M}}$ &\\
\noalign{\smallskip}\hline
\end{tabular}
%\end{table*}

\vspace{3mm}
{\noindent\bf Acknowledgment:} 
This research was supported by the German Federal Ministry of Education and Research (BMBF) under project reference 03FH025PX2 and Philips Lighting. We are indebted to our colleagues Ulrich Stein,  Thorsten Struckmann and Klaus Spohr for discussions. Help from Mads J. Herring Jensen concerning the pressure constraint is very much appreciated.

\pagebreak
\bibliographystyle{unsrt}
\bibliography{HIDLiteratur}   % name your BibTeX data base

\end{document}